\documentclass[longauth]{aa} 
\usepackage{graphicx}
\usepackage{txfonts}
\usepackage{natbib}
\usepackage{fixltx2e}
\usepackage{color}
\begin{document}
   \title{Discovery of VHE $\gamma$-rays from the blazar 1ES~1215+303 with the MAGIC Telescopes and simultaneous multi-wavelength observations}

   \author{ J.~Aleksi\'c\inst{1} \and
 E.~A.~Alvarez\inst{2} \and
 L.~A.~Antonelli\inst{3} \and
 P.~Antoranz\inst{4} \and
 S.~Ansoldi\inst{15} \and
 M.~Asensio\inst{2} \and
 M.~Backes\inst{5} \and
 U.~Barres de Almeida\inst{6} \and
 J.~A.~Barrio\inst{2} \and
 D.~Bastieri\inst{7}
% ...
}

\institute { IFAE, Edifici Cn., Campus UAB, E-08193 Bellaterra, Spain
 \and Universidad Complutense, E-28040 Madrid, Spain
 \and INAF National Institute for Astrophysics, I-00136 Rome, Italy
 \and Universit\`a  di Siena, and INFN Pisa, I-53100 Siena, Italy
 \and Technische Universit\"at Dortmund, D-44221 Dortmund, Germany
 \and Max-Planck-Institut f\"ur Physik, D-80805 M\"unchen, Germany
 \and Universit\`a di Padova and INFN, I-35131 Padova, Italy
 \and Inst. de Astrof\'{\i}sica de Canarias, E-38200 La Laguna, Tenerife, Spain
 \and University of \L\'od\'z, PL-90236 Lodz, Poland
 \and Depto. de Astrof\'{\i}sica, Universidad de La Laguna, E-38206 La Laguna, Spain
 \and Deutsches Elektronen-Synchrotron (DESY), D-15738 Zeuthen, Germany
 \and ETH Zurich, CH-8093 Zurich, Switzerland
 \and Universit\"at W\"urzburg, D-97074 W\"urzburg, Germany
 \and Universitat de Barcelona (ICC/IEEC), E-08028 Barcelona, Spain
 \and Universit\`a di Udine, and INFN Trieste, I-33100 Udine, Italy
% ...
}
   \date{}
\abstract{We present the discovery of very high energy (VHE,
E\,$>\,100$\,GeV) $\gamma$-ray emission from the BL Lac object
1ES~1215+303 by the MAGIC telescopes and simultaneous multi-wavelength
data in a broad energy range from radio to $\gamma$-rays.}
{We study the VHE $\gamma$-ray emission from 1ES~1215+303 and its
relation to the emissions in other wavelengths.}
{Triggered by an optical outburst, MAGIC observed the source in
2011 January-February for 20.3 hrs. The target
was monitored in the optical R-band by the KVA telescope that also
performed optical polarization measurements. We triggered target of
opportunity
observations with the \textit{Swift} satellite and obtained
simultaneous and quasi-simultaneous data from the \textit{Fermi} Large
Area Telescope and from the Mets\"ahovi radio telescope. We also
present the
analysis of older MAGIC data taken in 2010.}
{The MAGIC observations of 1ES~1215+303 carried out in
2011 January-February resulted in the first detection of the source at VHE
with a statistical significance of 9.4\,$\sigma$. Simultaneously, the
source was observed in a high optical and X-ray state. In 2010 the
source was
observed in a lower state in optical, X-ray, and VHE, while the GeV
$\gamma$-ray flux and the radio flux were comparable in 2010 and 2011.
The spectral
energy distribution obtained with the 2011 data can be modeled with a
simple one zone SSC model, but it requires extreme values for the
Doppler factor or the electron energy distribution.}{} 
 
   \keywords{Gamma rays:galaxies--BL Lacertae objects:individual:
1ES~1215+303}
\titlerunning{1ES~1215+303}
   \maketitle
%

\section{Introduction}

Most of the extragalactic sources from which very high energy (VHE,
$>$100\,GeV) $\gamma$-ray emissions has been detected are
blazars. These objects are commonly believed to be a subtype of active
galactic nuclei (AGN) whose relativistic jet points very close to the
line of sight of the observer. Blazars are characterized by high
amplitude 
variability at all wavebands from radio to $\gamma$-rays.  The
correlations between the different energy bands are complicated, but
in general it seems that high states in lower energy bands
(e.g. optical) are accompanied by high states in the higher energies
(i.e. $\gamma$-rays) at least in some sources \citep[see
e.g.][]{2012JPhCS.355a2013R}.

The spectral energy distribution (SED) of blazars exhibits a generic
two-bump structure: one peak with a maximum in the spectral range from
 radio to X-rays  and
a second peak in the interval from X-ray to $\gamma$-ray. The radiation
is
produced in a highly beamed plasma jet and the double peaked SED is
 often explained by a single population of relativistic electrons. The
first peak is due to synchrotron emission in the magnetic field of the
jet and the second peak is caused by inverse Compton (IC) scattering
of low-energy photons \citep{1967MNRAS.137..429R}. The low-energy
photons can originate externally to the
jet \citep[external Compton scattering,][]{1993ApJ...416..458D} or be
produced within the jet
via synchrotron radiation \citep[synchrotron self-Compton scattering,
SSC,][]{1992ApJ...397L...5M}.

Blazar is a common term used for Flat Spectrum Radio Quasars (FSRQs)
and BL~Lac objects (BL Lacs), which are thought to be intrinsically
different. The FSRQs show broad emission lines in their optical
spectra while the BL Lacs have featureless spectra with weak or no
emission lines possibly masked by a strong emission from the jet. This
indicates that in BL Lac objects the main
population of seed photons for Compton scattering should originate
from the synchrotron emission. Indeed most of the SEDs of BL Lacs are
well described with simple SSC model
\citep[e.g.][]{1996ApJ...461..657B,1998ApJ...509..608T}.

MAGIC has been successfully performing optically triggered VHE
$\gamma$-ray observations of AGN since the start of its
operations. The triggers have been provided by the Tuorla blazar
monitoring program{\footnote{http://users.utu.fi/kani/1m/}} and the
target of opportunity (ToO) observations with MAGIC have 
resulted in the discovery of five new VHE $\gamma$-ray emitting
sources (Mrk~180, \cite{2006ApJ...648L.105A}; 1ES~1011+496,
\cite{2007ApJ...667L..21A}; S5~0716+714, \cite{2009ApJ...704L.129A};
B3~2247+381, \cite{2012A&A...539A.118A}; and 1ES~1215+303, this
paper). However, in many cases it has not been possible to confirm
with high statistical significance if the sources were in higher VHE
$\gamma$-ray state than usual during the observations.  The long-term
studies
of individual VHE $\gamma$-ray blazars like Mrk~421
\citep{2011ApJ...738...25A} and PG~1553+113 \citep{2012ApJ...748...46A}
have also yielded controversial results on  the
correlation between the two broad energy ranges.  Thus, to date, the
connection
between the optical and VHE $\gamma$-ray states has remained an open
question.

1ES~1215+303 (also known as ON~325) is a high
synchrotron peaking BL~Lac object \citep{2010ApJ...722..520A} with
redshift $z=0.130$ \citep[however, 
$z=~0.237$ is also reported in the literature,
e.g.
NED\footnote{http://nedwww.ipac.caltech.edu/}]{2003ApJS..148..275A}.
The source was classified as a promising candidate TeV blazar
\citep{2002A&A...384...56C,2010MNRAS.401.1570T} and has been observed
several times in VHE $\gamma$-rays before the observations presented
here. The previous observations yielded only upper limits,
Whipple: F\,$(>430\mathrm{GeV})<1.89\times10^{-11}$~cm$^{-2}$ s$^{-1}$,
\citep{2004ApJ...603...51H}; MAGIC:
F\,$(>120\mathrm{GeV})<3.5\times10^{-11}$~cm$^{-2}$ s$^{-1}$,
\citep{2011ApJ...729..115A}. The source was listed in the {\it Fermi}
Large Area Telescope (LAT)
bright AGN catalog \citep{2009ApJ...700..597A} as showing a hard
spectrum
($\Gamma=1.89\pm0.06$).  It underwent a large outburst 
in late 2008, and in this catalog
1ES~1215+303 is the only high energy peaking source that shows
significant variability. In the second {\it Fermi}-LAT AGN catalog
\citep{2011ApJ...743..171A} other high synchrotron peaking sources
have also been flagged as variable.

In the first days of 2011 January 1ES~1215+303 was observed to be in a
high optical state. This triggered MAGIC observations, extending until 2011
February, that 
resulted in the discovery of VHE $\gamma$-rays from the source
\citep{2011ATel.3100....1M}. In this paper we present the results of
the 2011 January-February observations. We also present the previous
observations of 1ES~1215+303 with the MAGIC telescopes performed in
2010 January-February and 2010 May-June that produced only a hint
of signal. For all epochs we present simultaneous and
quasi-simultaneous multi-wavelength data from radio, optical, X-ray,
and GeV $\gamma$-rays.

\section{Observations and Data Analysis}
The observations of 1ES~1215+303 were performed in a broad wavelength
range (from radio to VHE $\gamma$-rays) by 5 different
instruments. This is the first time that such a broad wavelength range
is covered for this source in quasi-simultaneous observations.

\subsection{MAGIC}

MAGIC consists of two 17~m Imaging Air Cherenkov Telescopes
(IACTs) sensitive to $\gamma$-rays with energy above 50\,GeV in
standard trigger mode (which is the
lowest trigger energy threshold among the existing IACTs). The system
is located in the Canary Island of La Palma, 2200~m above sea
level. Since fall 2009 the telescopes are working together 
in stereoscopic mode which ensures a sensitivity of $<0.8\%$ of the 
Crab Nebula flux above 300\,GeV in 50~hrs of observations
\citep{2012APh....35..435A}. 
The field of view of the each MAGIC camera has a diameter of
3.5$^\circ$.

1ES~1215+303 was observed by MAGIC in 2010 January-February, 2010 May-June
, and 2011 January-February, for a total of 48 hrs. The
observations were done in the so-called wobble mode (i.e. with the
source offset by $0.4^\circ$ from the camera center), which provides a
simultaneous estimate of the background from the same data set
\citep{1994APh.....2..151F}. While most of the data were taken in dark
night conditions, a small fraction were taken in presence of moderate
moonlight. The data span a range of zenith angle from $1^\circ$ to
$40^\circ$ with most of the data taken below $25^\circ$ (in 2010 the
mean zenith angle was $\sim13^\circ$, and in 2011 $\sim8^\circ$).

The data were analyzed using the standard MAGIC software and analysis
package \citep{2012APh....35..435A}.  
Another VHE $\gamma$-ray emitter,
1ES~1218+304 \citep{2006ApJ...642L.119A} is present in the same
field of view as 1ES~1215+303. The sources are separated by $\sim
0.8^\circ$, which is much larger than the point spread function
(PSF) of the MAGIC telescopes ($\sim 0.1^\circ$), so there was no
source confusion or contamination.
However, these sources have nearly the
same Right Ascension, so in the standard wobble setup used in the
2010 January-February observations, the background estimation region
partially overlapped with the 1ES~1218+304 position. This would
result in an overestimate of the background, so this region was
excluded from the background estimate. In the later observations
(2010 May-June, and 2011 January-February) the wobbling offset
direction was changed to have the standard background
estimation regions far from the second source.

After the data quality selection, based mainly on the rate of stereo
events,
the data samples of January-February 2010, May-June 2010
and January-February 2011  contain 19.4, 3.5, and 20.6~hrs of good
quality data respectively. Because of the 
different positions of the source in the camera, 
and the variable nature of AGN,
we decided to split the analysis into these 3 periods.

\subsection{Fermi-LAT}
The {\it Fermi}-LAT is a pair conversion telescope designed to cover
the energy band from 20\,MeV to greater than 300\,GeV. It operates in
all-sky survey mode and therefore can provide observations of
1ES~1215+303 simultaneous to MAGIC. In this paper 
the standard LAT Science Tools (version v9r23p0) were used 
to analyze the data collected in
the time interval from 2008 August 5 to 2011 March 22. For this
analysis, only events belonging to the ``Diffuse'' class (which have
the highest probability of being photons) and located in a
circular Region Of Interest (ROI) of $7^{\circ}$ radius, centered at
the
position of 1ES~1215+303, were selected (using Pass 6 event selection).
In addition, we applied a cut on the zenith angle ($< 100^\circ$) limb
$\gamma$-rays and
a cut on the rocking angle ($> 52^\circ$) to limit Earth limb
contamination.

The data analysis of 1ES~1215+303 is very challenging due to the
presence 
of several $\gamma$-ray emitting sources in the same ROI.  
1ES~1218+304 is located at a distance of just $0.8^\circ$ from
the source of interest. Another well known VHE emitter,
W~Comae, is located at $\sim2^\circ$ from the latter source. Thus, the
LAT analysis was restricted to energies above 1\,GeV
where the {\it Fermi}-LAT PSF is sufficiently
narrow{\footnote{http://www.slac.stanford.edu/exp/glast/groups/canda/archive/pass6v11\\/lat\_Performance.htm}}
to separate 1ES~1215+303
from the other sources in the ROI. The unbinned likelihood method was
applied to
events in the energy range from 1\,GeV to 300\,GeV. All
point sources from the 2FGL~catalog \citep{2012ApJS..199...31N}
located within $12^\circ$ of 1ES~1215+303 were included in the model of
the
region. Sources located within a $5^\circ$ radius centered on
1ES~1215+303 position had their flux and photon index left as free
parameters. The diffuse Galactic and isotropic components (including
residual instrumental background) were modeled with the publicly
available files {\tt gll\_iem\_v02\_P6\_V11\_DIFFUSE.fit} and
isotropic {\tt iem\_v02\_P6\_V11\_DIFFUSE.txt}
{\footnote{http://fermi.gsfc.nasa.gov/ssc/data/access/lat/BackgroundModels.html}}. The
normalizations of the components comprising the total background model
were allowed to vary freely during the spectral point fitting. The
instrument response functions {\tt P6\_V11\_DIFFUSE} were used. The
successful separation of flux between 1ES~1215+303 and 1ES~1218+304
was verified by the absence of any significant correlation between
their light curves.  The systematic uncertainty in the flux is
estimated as 
$5\%$ at 560\,MeV and $20\%$ at 10\,GeV and above {\footnote{http://fermi.gsfc.nasa.gov/ssc/data/analysis/LAT\_caveats.html}}.

\subsection{Swift}  

The \textit{Swift} satellite \citep{2004ApJ...611.1005G} is equipped
with
three telescopes, the Burst Alert Telescope
\citep[BAT;][]{2005SSRv..120..143B} covering the 15-150\,keV
energy range, the X-ray telescope \citep[XRT;][]{2005SSRv..120..165B}
covering the 0.2-10\,keV energy band, and the UV/Optical Telescope
\citep[UVOT;][]{2005SSRv..120...95R} covering the 180--600\,nm
wavelength range.

A \textit{Swift} ToO request was submitted on 2011 January
3. The \textit{Swift} observations started on January 4
until January 12 with four $\sim$ 5\,ks exposures in photon counting
mode. The data were processed with standard procedures using the
FTOOLS task XRTPIPELINE (version 0.12.6) distributed by HEASARC within
the HEASoft package (v.6.10). Events with grades 0--12 were selected
for the data \citep[see][]{2005SSRv..120..165B}
and the latest response matrices available in the \textit{Swift}
CALDB (v.20100802) were used. For the spectral analysis, we extracted
the source events in the 0.3-10\,keV range within a circle with a
radius of 20 pixels ($\sim 47\,$ arcsec). The background was extracted
from an off-source circular region with a radius of 40 pixels.

The spectra were extracted from the corresponding event files and
binned using GRPPHA to ensure a minimum of 25 counts per energy bin, in
order to obtain reliable $\chi^2$ statistics.
Spectral analysis was performed using XSPEC version 12.6.0.
The neutral  hydrogen-equivalent column density was fixed to the
Galactic value in the direction of the source $1.74\cdot
10^{20}$~cm$^{-2}$ \citep{2005A&A...440..775K}.

{\it Swift}$/$UVOT observed the source with all filters (V, B, U, UVW1,
UVM2, UVW2) for four nights. UVOT source counts were extracted from a
circular region 5 arcsec-sized centered on the source position, while
the background was extracted from a larger circular nearby source--free
region. These data were processed with the {\tt uvotmaghist} task of
the HEASOFT package.
The observed magnitudes have been corrected for Galactic extinction
$E_{B-V}=0.024$ mag \citep{1998ApJ...500..525S}
, applying the formulae by \cite{1992ApJ...395..130P} 
 and finally converted into fluxes following
\cite{2008MNRAS.383..627P}.  

\subsection{KVA}

The KVA optical telescopes are located in La Palma, but are operated
remotely from Finland. The two telescopes are attached to the same
fork. The larger telescope has a mirror diameter of 60\,cm and the
smaller
35\,cm.  

The 35\,cm telescope is used for simultaneous photometric observations
with
MAGIC, but also to monitor potential VHE $\gamma$-ray candidate AGN in
order to trigger MAGIC observations if the sources are in high optical
states. The observations are performed in the R-band and the
magnitude of the source is measured from CCD images using
differential photometry. The comparison star magnitudes for
1ES~1215+303 are from \cite{1996A&AS..116..403F}, and the magnitudes
are converted to flux using the formula and values from
\cite{1979PASP...91..589B}.
1ES~1215+303 has been observed regularly as part of the Tuorla
blazar monitoring program since 2002. 

The 60\,cm telescope is used for polarimetric observations \citep[see
e.g.][]{2005ApJ...632..576P, 2011A&A...530A...4A}. For 1ES~1215+303
polarimetric
observations were performed on six nights from 2011 January 7 to January 17.
 The 
degree of polarization and position angle were calculated from the
intensity
ratios of the ordinary and extraordinary beams using standard formulae
and semiautomatic software specially developed for polarization
monitoring purposes. 

\subsection{Mets\"ahovi radio telescope}
37\,GHz radio observations were made with the 13.7\,m Mets\"ahovi
radio telescope located in Kylm\"al\"a, Finland. The telescope, the
observation methods, and the data analysis procedure are described in
e.g. \cite{1998A&AS..132..305T}. The telescope detection limit is
$\sim$ 0.2~Jy under optimal conditions and since 1ES~1215+303 is
a rather weak source at 37\,GHz it can only be observed under good
weather conditions. Typically, an acceptable measurement of the source
is obtained approximately once per month. Data were obtained
simultaneously with the MAGIC
observations in 2010 June, but in 2011 January-February the weather did not
allow simultaneous observations with MAGIC, the
closest points being from 2010 December and 2011 March.

\section{Results}

\subsection{MAGIC results}

\begin{figure*}
\includegraphics[scale=0.33, trim= 10 15 25 23,
clip]{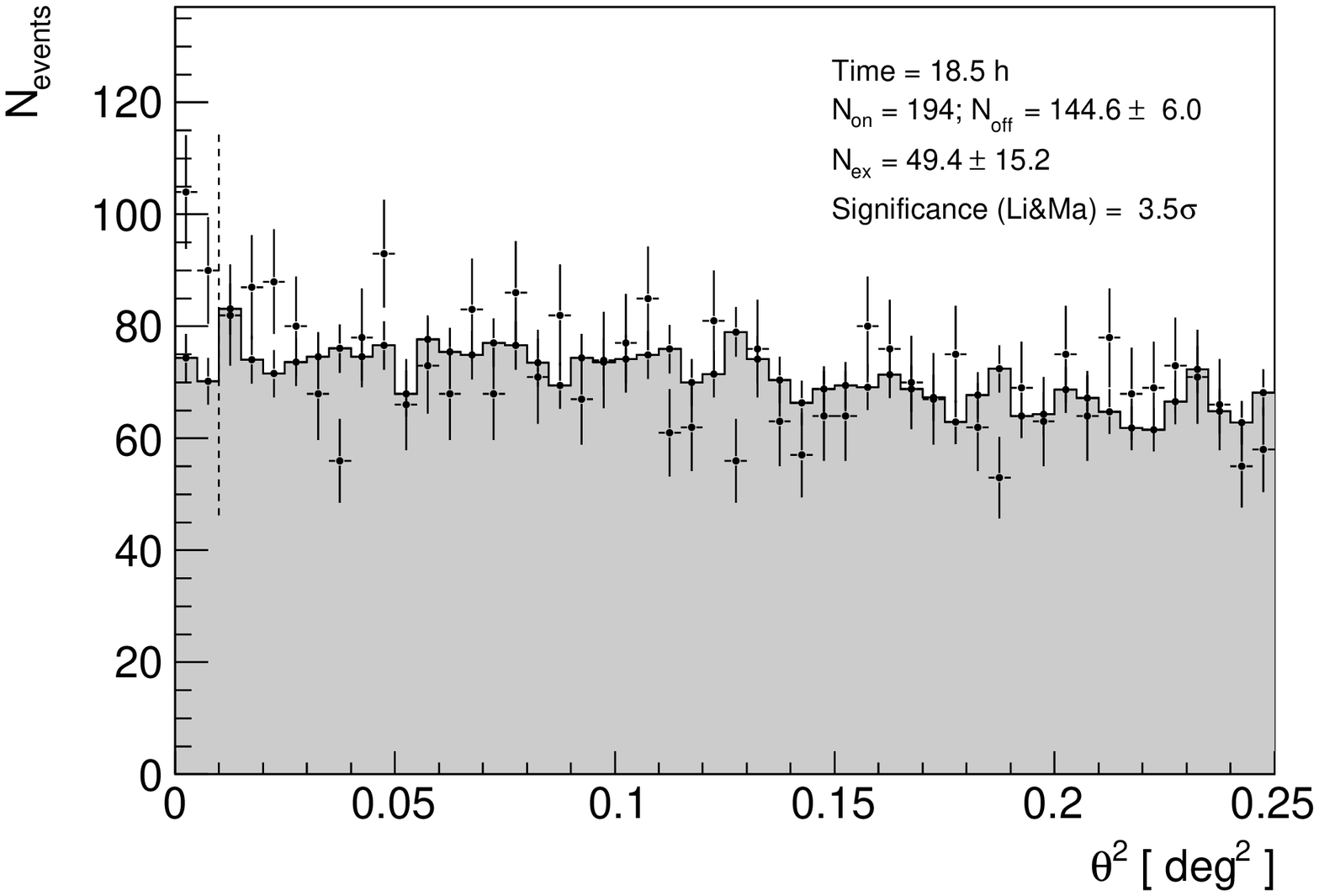}
\includegraphics[scale=0.33, trim= 10 15 25 23,
clip]{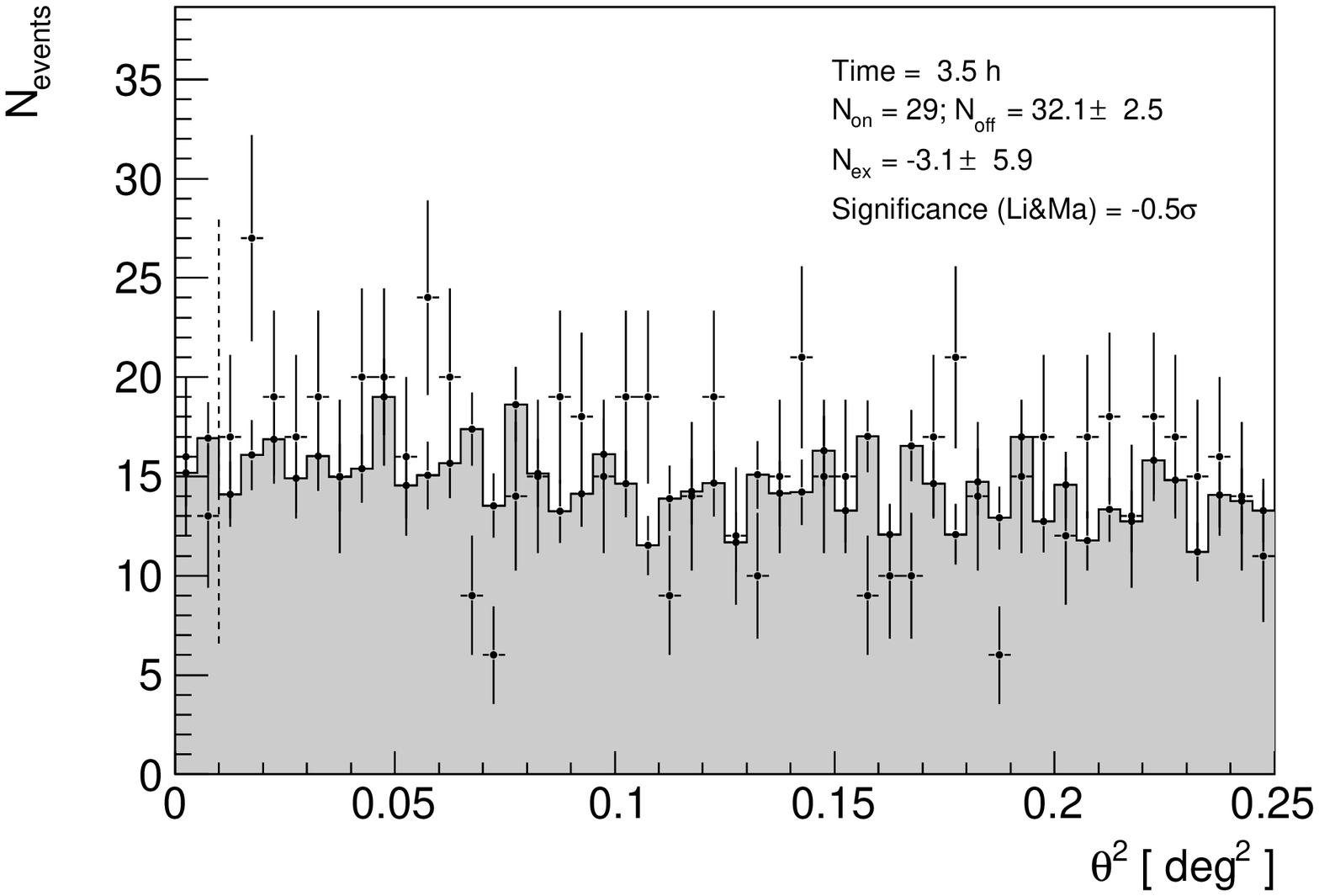}
\includegraphics[scale=0.33, trim= 10 15 25 23,
clip]{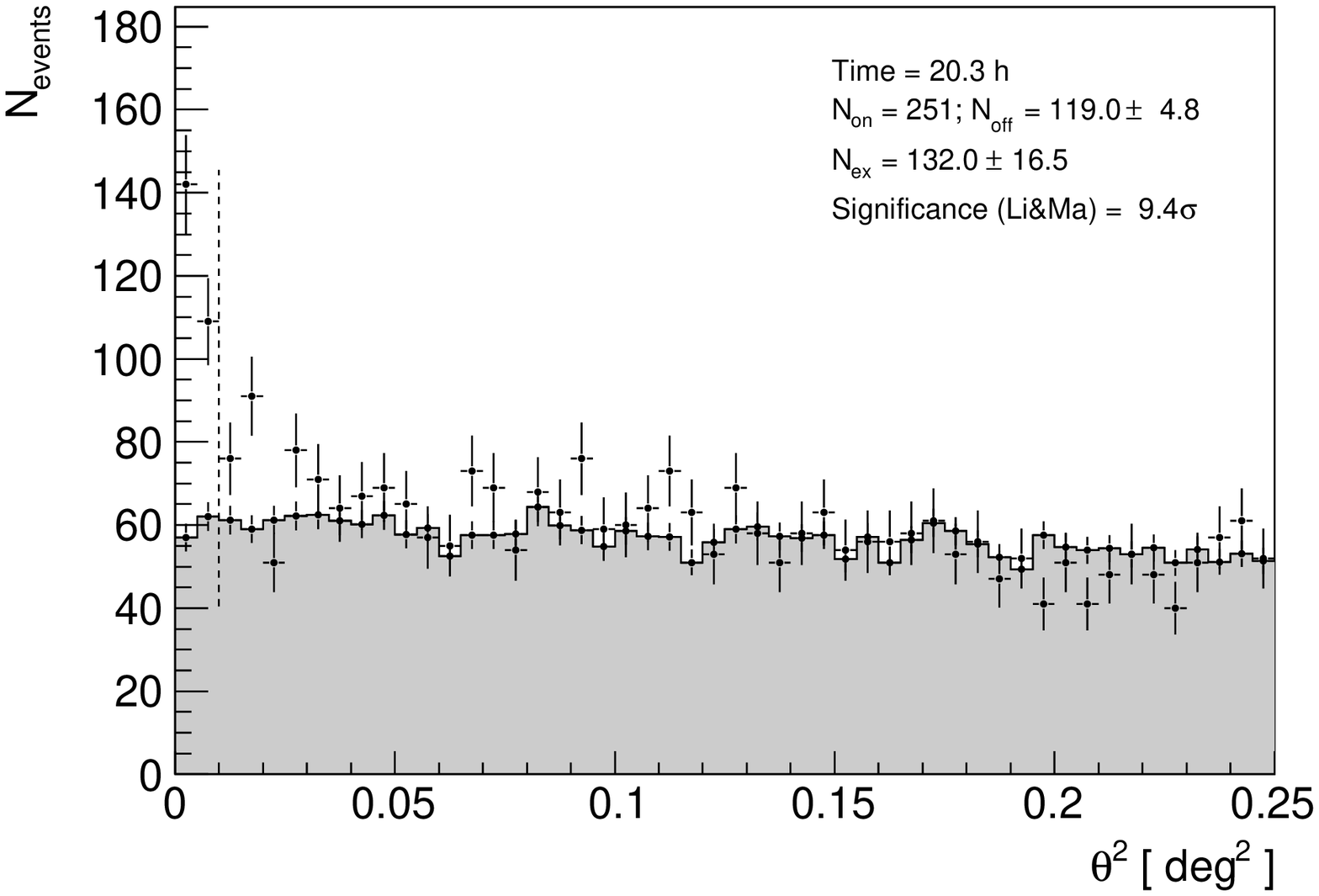}
\caption{Distributions of the $\theta^2$ parameter for 1ES~1215+303
signal (black histograms) and background estimation (gray histograms)
for the three observation periods: January-February 2010 (left),
May-June 2010 (middle), and January-February 2011 (right). The vertical
dashed line corresponds to the apriori defined signal region
$\theta^2<0.01$ deg$^2$.}
\label{fig_theta2}
\end{figure*}

The MAGIC data were divided in three samples corresponding to three
observation epochs: 2010 January-February, 2010 May-June, and 2011
January-February.
The so-called $\theta^2$ plots (the distribution of the squared
angular distance between the arrival direction of the events and the
real position), 
for energies above 300\,GeV, corresponding to
the three observation epochs, are shown in Fig.~\ref{fig_theta2}. The
computation of the number of the ON (signal) and OFF (background)
events was performed in a fiducial signal region of
$\theta^2<0.01\,\mathrm{deg}^2$, and using 5 background regions (4 in
case of the January-February 2010 data). In 2010 January-February (left
panel) 194 ON events were detected over $144.6\pm6.0$ OFF events, with
a significance level of
3.5$\,\sigma$ \citep[using Eq. 17 in][]{1983ApJ...272..317L}.
In 2010 May-June (middle panel) the observation time was much shorter
and no excess events were present. For 2011 data (right panel) the
numbers were 251 ON over $119\pm4.8$ OFF corresponding to a 
$\sim$9.4\,$\sigma$ significance, which is the first detection of VHE
$\gamma$-rays from this source.

\begin{figure}
\includegraphics[width=0.5\textwidth]{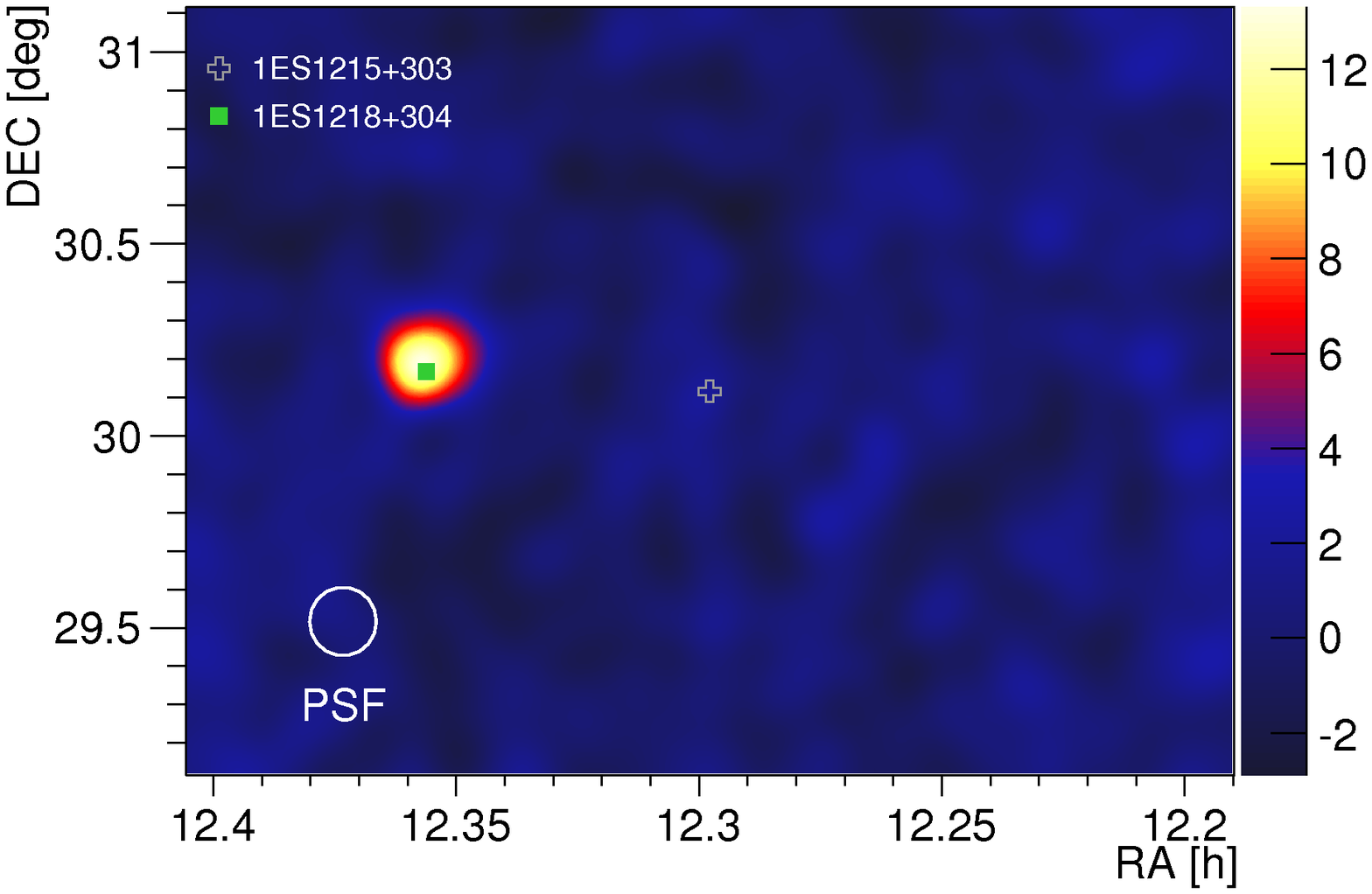}
\includegraphics[width=0.5\textwidth]{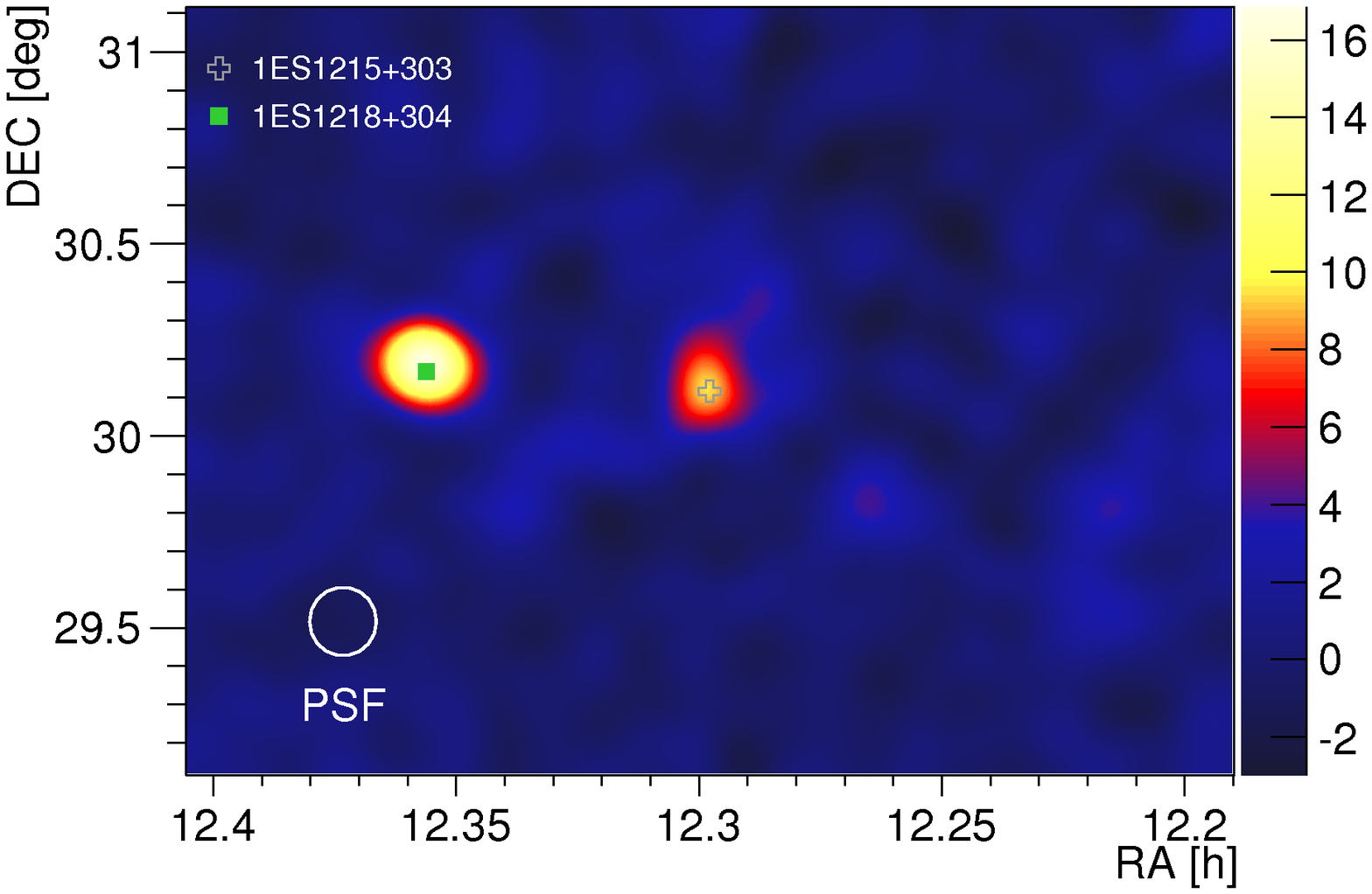}
\caption{Significance maps ($>300$\,GeV) from MAGIC observations
performed during 2010 January-February and 2010 May-June (combined
together, total time 22.0 hrs, top) and 2011 January-February (total
time 20.3 hrs, bottom).}
\label{fig_skymap}
\end{figure}

In Fig.~\ref{fig_skymap}, we show the significance map of the sky
region for energies above
300\,GeV for the 2010 (January-February and May-June combined) and 2011
observations. 1ES~1218+304 is clearly visible in both maps while
1ES~1215+303 was fainter in 2010 than in 2011. The 1ES~1218+304 data
analysis and results will be addressed in a separate paper.

For the light curve and spectrum determinations softer cuts were applied
 that have a higher
$\gamma$-ray efficiency. The light curve (in a 5-days bins)
above 200\,GeV of the 2011 data is well described by a constant flux
of $(7.7 \pm 0.9) \times 10^{-12}\,\mathrm{cm^{-2} s^{-1}}$
($\chi^2/n_{\rm dof}$ = 0.56 / 3), which corresponds to about 3.5\% of the
Crab Nebula flux. Assuming that the hint of a signal seen in the 2010
data is a $\gamma$-ray excess the corresponding flux was
F\,$(>200$\,GeV)$=(3.4 \pm 1.0) \times
10^{-12}\,\mathrm{cm^{-2} s^{-1}}$, which is less than half of the flux
measured in 2011.
The hypothesis of constant flux between 2010 and 2011 is excluded at
the level of 3.1\,$\sigma$. 

\begin{figure}
\includegraphics[width=0.47\textwidth]{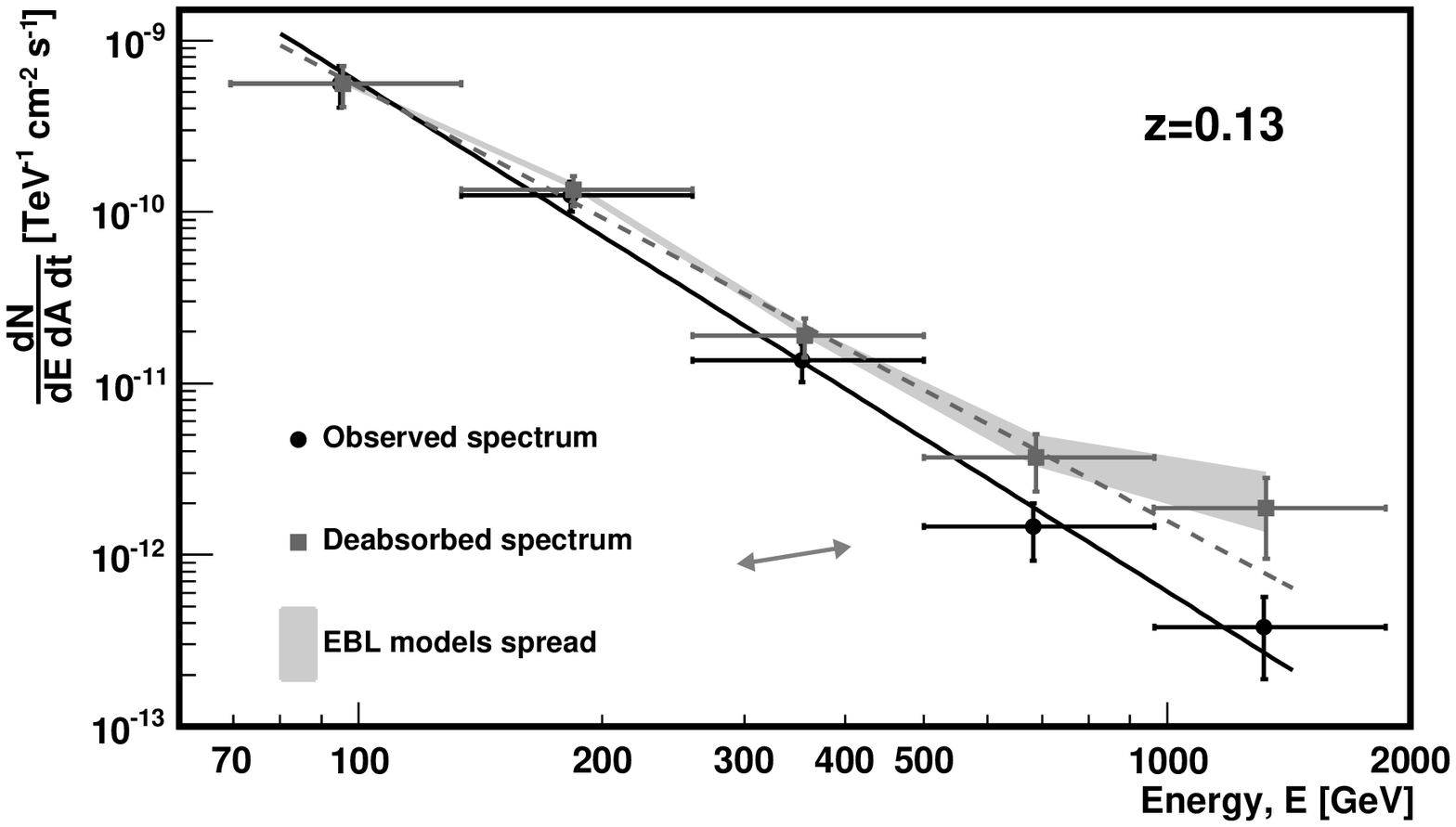}
\caption{Observed and deabsorped VHE $\gamma$-ray spectra for a
redshift of 0.130. The EBL model of \cite{2011MNRAS.410.2556D} was
used, the gray area shows the spread of the EBL models. The arrow
shows the systematic error of the measurement.}
\label{fig_spec}
\end{figure}

The derived VHE $\gamma$-ray spectrum for the 
2011 observations can be described by a single power law ($\chi^2 /
n_{\rm dof} = 5.2 / 3$, see Fig.~\ref{fig_spec}) : 
\begin{equation}
\frac{\mathrm{d}N}{\mathrm{d}E} = (2.27\pm
0.25)\times10^{-11}\,\Big(\frac{E}{300\mathrm{GeV}}\Big)^{(-2.96\pm
0.14)}\,\mathrm{TeV}^{-1} \mathrm{cm}^{-2} \mathrm{s}^{-1}
\end{equation}
in the fitting range 70\,GeV -- 1.8\,TeV. Since the
spectral index of 1ES~1215+303 is similar to that of the Crab
Nebula and the source is relatively bright, we can directly use the
systematic errors estimated in \cite{2012APh....35..435A}. 
The systematic error of the slope is $\pm0.15$ and in the energy range
of the 1ES~1215+303 spectrum, the error in the
flux normalization (without the energy scale uncertainty) was estimated
to be 11\%. 
The systematic error in the energy scale is 15\%. Finally,
the MAGIC spectrum was deabsorbed using different EBL models
\citep{2011MNRAS.410.2556D,2010A&A...515A..19K,2008A&A...487..837F,2005AIPC..745...23P}
and the maximum high UV EBL model described in
\cite{2008Sci...320.1752M} for $z=0.130$. The results are shown in
Fig.~\ref{fig_spec}. As denoted in the Figure by the shaded area, at
this redshift the EBL models agree well. The EBL model of
\cite{2011MNRAS.410.2556D} was used to calculate the final intrinsic
spectrum since this model is based on an observational approach.


\subsection{Fermi-LAT results}

\begin{figure}
\includegraphics[width=0.18\textwidth,
angle=270]{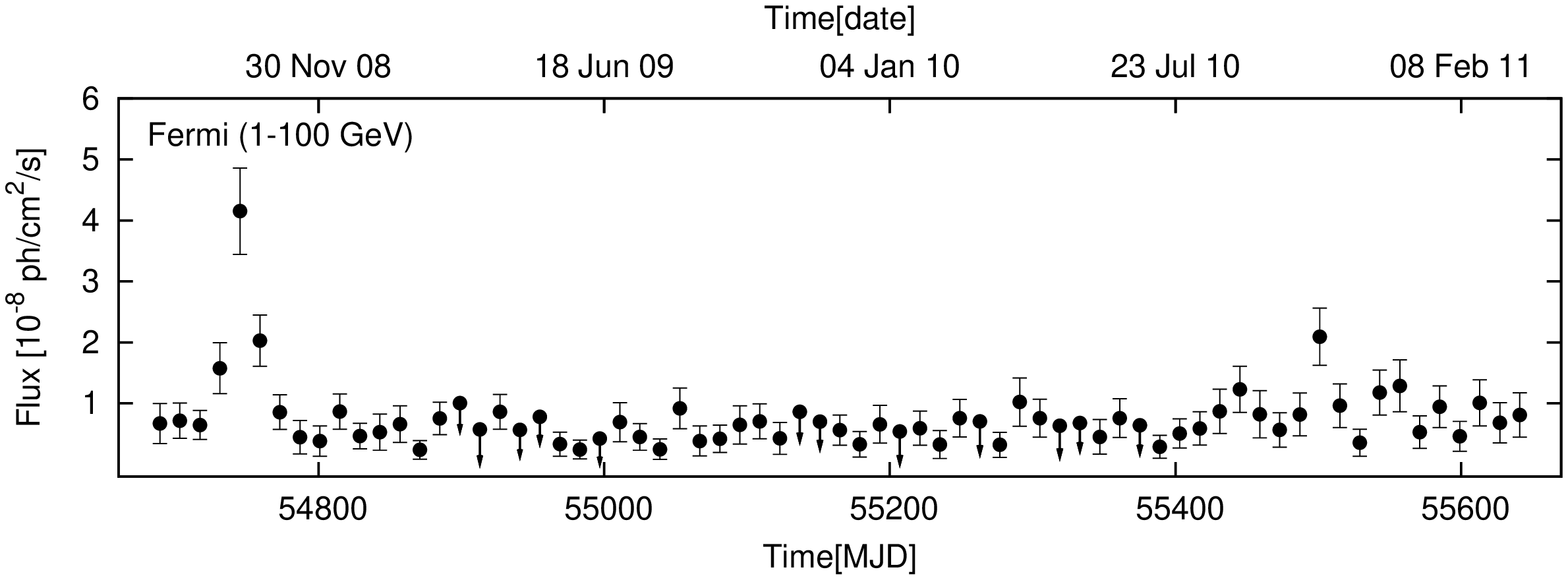}
\caption{Long-term light curve of 1ES~1215+303 from {\it Fermi}-LAT
between 1 and 100\,GeV.} 
\label{fig_fermilc}
\end{figure}

\begin{figure}
\includegraphics[width=0.49\textwidth]{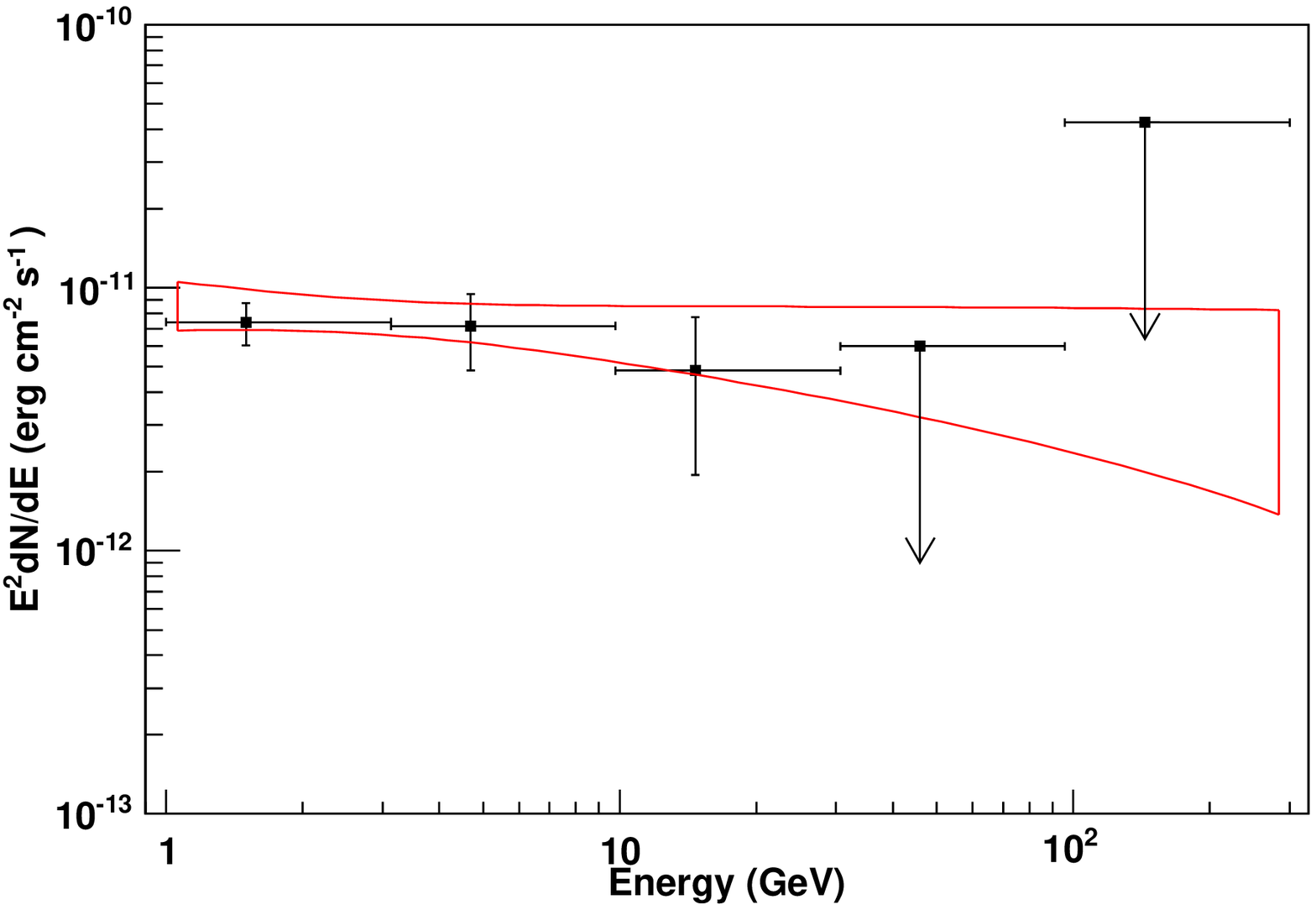}
\includegraphics[width=0.49\textwidth]{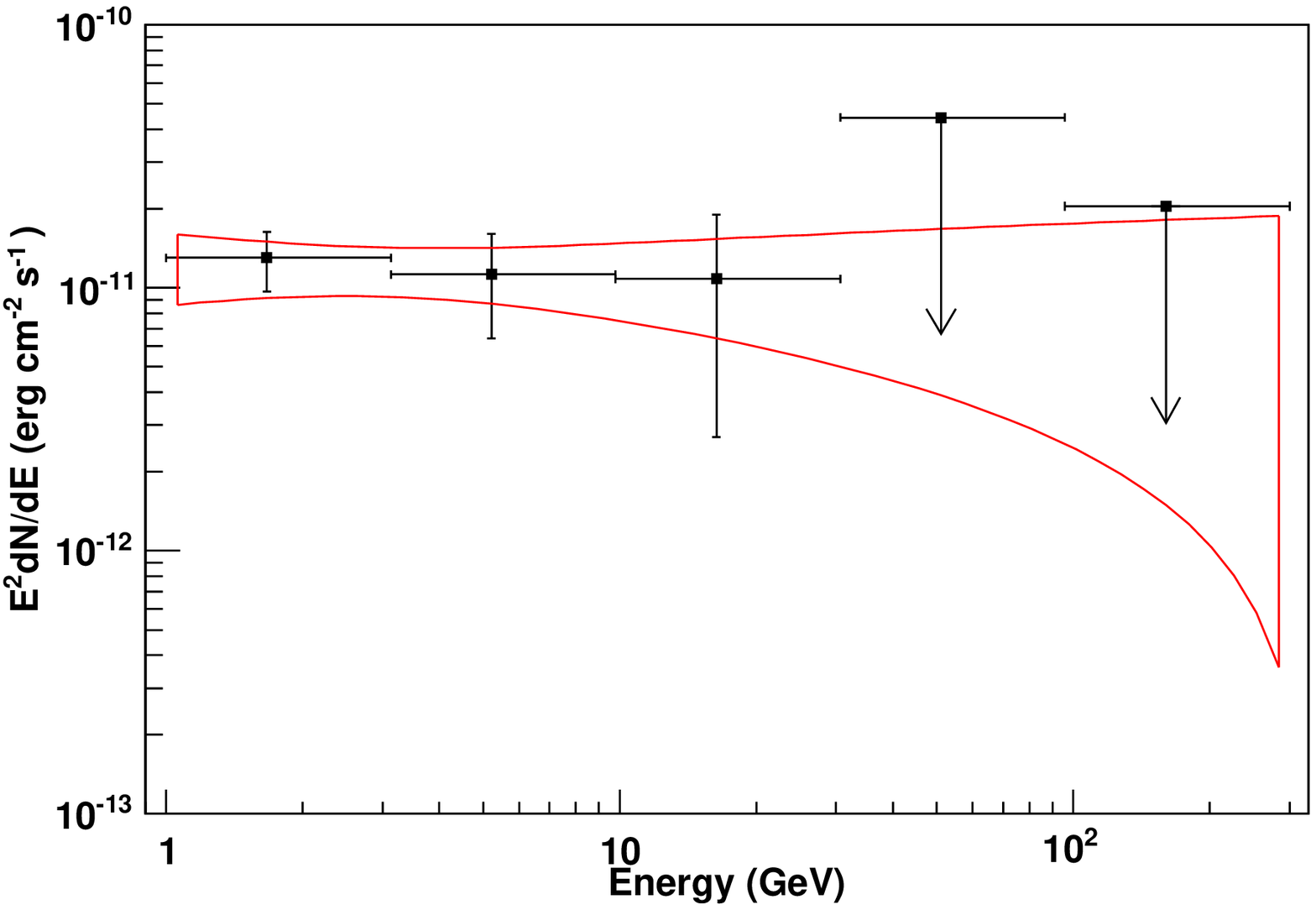}
\caption{Spectral energy distribution from {\it Fermi}-LAT derived for
2011 January-June (top) and 2011 January-February (bottom). The upper
limits have been computed when the test statistics \citep[see
e.g.][]{1996ApJ...461..396M} in the energy band were lower than 4. The
bow-ties are derived from the unbinned likelihood analysis.} 
\label{fig_fermised}
\end{figure}

The light curve of 1ES~1215+303 was obtained in the energy range from 1
to 100\,GeV, in
14--day bins from 2008 August to 2011 March (Fig.~\ref{fig_fermilc}). It shows the major flare reported in
\cite{2009ApJ...700..597A} at the beginning of the Fermi
mission. There is a hint of enhanced flux during 2010 November (MJD
55500,
duration only one bin, i.e. 14 days) but very little variability
otherwise, especially at the two MAGIC observation epochs
(2010 January-June and 2011 January-February). To maximize the number
of photons the spectral energy distribution was derived using the whole MAGIC dataset 
(2010 January-June and 2011 January-February). The spectral energy
distributions are shown in Fig.~\ref{fig_fermised}. In 2010 January-June
the integral flux, F\,$(1-100$\,GeV) is
$(4.9\pm0.7)\times10^{-9}$\,cm$^{-2}$s$^{-1}$ and the photon index
$2.1\pm0.1$, while in 2011 January-February F\,($1-100$\,GeV)$=
(7.3\pm1.6)\times10^{-9}$\,cm$^{-2}$s$^{-1}$ and the photon index
$2.0\pm0.2$. The mean detected flux was $\sim50\%$ higher in
2011 January-February than in 2010 January-June, but 
due to large error bars the increase was not statistically significant.
The spectral index was constant within the error bars.

\subsection{Swift results}
The results of the {\it Swift}/XRT observations are summarized in
Table~1. The source showed the highest flux on 2011 January 8 (MJD
55569.1) and previous/subsequent observations from 2009 December (MJD
55168.7)/ 2011 April (MJD 55674.2) show significantly lower flux. For
the X-ray spectra both log parabola (in the form $\sim
E^{-a-b*{\mathrm log}(E)}$, with $E$ being the energy in keV) and a
simple power-law fit
were tested. The best fit was achieved with a log parabola law model
in the
range 0.3--10 keV for four observations while a
simple power law, in the range 0.5--10 keV, provided a better fit for
three of the observations. Generally, a log parabolic fit suggests that
there is curvature in the X-ray spectra but for
1ES~1215+303, the difference between log parabolic and power law fits
is small so no strong conclusions can be drawn. Because of the different
fits a comparison between the spectral slopes is difficult, but for the
highest flux night the spectral index is marginally harder than for the
low state observations.

\begin{table*}[!h]
\begin{center}
\begin{tabular}{l|l|l|l|l|l}
\hline
\hline
MJD & Obs. Time [ks] &  Flux (2-10\,keV) [$10^{-12}$ erg/cm$^2$/s] &
 $a$ ($\Gamma$ for PL) & $b$ &  $\chi^2_{\mathrm red}$/ $n_{\rm dof}$
\\ \hline 
55168.6799 &   4.99      &  $1.21 \pm  0.19$  &   $2.56 \pm  0.10$  &
 $0.34 \pm 0.34$ & 1.19/25 \\
55565.0340 &    4.39     &  $2.74 \pm  0.25$  &   $2.41 \pm  0.08$  &
 $0.37 \pm 0.24$ & 0.87/42 \\
55569.1281 &    2.38     &  $3.02 \pm  0.40$  &   $2.29 \pm  0.16$  &
  -- & 1.23/18 (PL)\\
55571.1327 &    4.07     &  $1.69 \pm  0.17$  &   $2.65 \pm  0.14$  &
-- & 1.23/18 (PL) \\
55572.1361 &    4.27     &  $1.45 \pm  0.20$ &   $2.64 \pm  0.09$  &
$0.28 \pm 0.27$ & 1.15/32 \\
55573.1396 &    2.99     &  $1.73 \pm   0.25$ &   $2.46 \pm  0.11$  &
$0.66 \pm 0.37$ & 1.26/25 \\
55674.2438 &    2.34     &  $1.30 \pm  0.30$ &   $2.67 \pm  0.25$  &
 --  & 0.48/8 (PL) \\ \hline
\end{tabular}
\caption{Data summary and results for the {\it Swift}/XRT ToO
observations. The datasets in the first/last rows are prior/subsequent
to the MAGIC observations and are reported for comparison. For each
dataset the following quantities are reported: the MJD time of the
beginning of the observations; the exposure time; the integral flux in
the 2-10 keV band; the $a$ and $b$ parameters for the log parabola fit
(or the photon index $\Gamma$ in case a simple power-law is used, see
text); the reduced $\chi^2$ with number of degrees of freedom $n_{\rm dof}$. PL indicates when the simple power law is used instead of the log parabola.}
\end{center}
\end{table*}

The {\it Swift}/UVOT results from 2011 January ToO observations show
constant brightness with V-band magnitude $=15.06\pm 0.10$, B$=15.38\pm
0.10$, U$=14.53\pm 0.08$, UVW1=$14.43\pm 0.08$, UVM2=$14.35\pm 0.06$,
and
UVW2=$14.46\pm 0.06$. However, in all bands the source is clearly
brighter than in the previous observation (2009 December: V$=15.60\pm
0.10$, B$=15.95\pm 0.10$, U$=15.12\pm 0.08$, UVW1$=15.07\pm 0.08$,
UVM2$=15.00\pm 0.06$, and UVW2$=15.15\pm 0.06$).

\subsection{KVA and Mets\"ahovi results}

In the optical R-band the source is clearly variable on daily and
yearly time-scales. The host galaxy contributes a flux of 
$0.99\pm0.09$ mJy \citep{2007A&A...475..199N} and when
this contribution was subtracted from the measured flux, the AGN core
was found to be $\sim40\%$ brighter in 2011 January-February (average
total flux 3.64\,mJy) than in 2010 January-June (average total flux 2.55\,mJy). Similarly, it was found that during the 2011 January-February observations the flux varied by $\sim$25\% (core flux between 3.2\,mJy and 4.1\,mJy).  

During 2011 January the optical polarization was $\sim9\%$ while during
the follow up observation in 2011 April it was  higher, $\sim15\%$.
The position angle (PA) was only slightly variable between 140 and 150
degrees.

In the radio band the source is rather weak and does not show strong
variability. The 37\,GHz flux from the Mets\"ahovi radio telescope has
a similar level (0.3-0.4\,Jy) in 2010 and 2011, although there were 
no radio observations during 2011 January-February.

\section{Interpretation}

In this section we discuss the quasi-simultaneous light curves, showing
how they establish
connections between different energy regimes and locate the emission
region. The spectral energy distribution is reconstructed for the
first time from radio frequencies to TeV energies for 1ES~1215+303,
allowing us to study the capability of the one-zone synchrotron self
Compton model to reproduce the constructed SED.

\subsection{Multi-wavelength behavior}

\begin{figure}
\includegraphics[width=0.55\textwidth, angle=270]{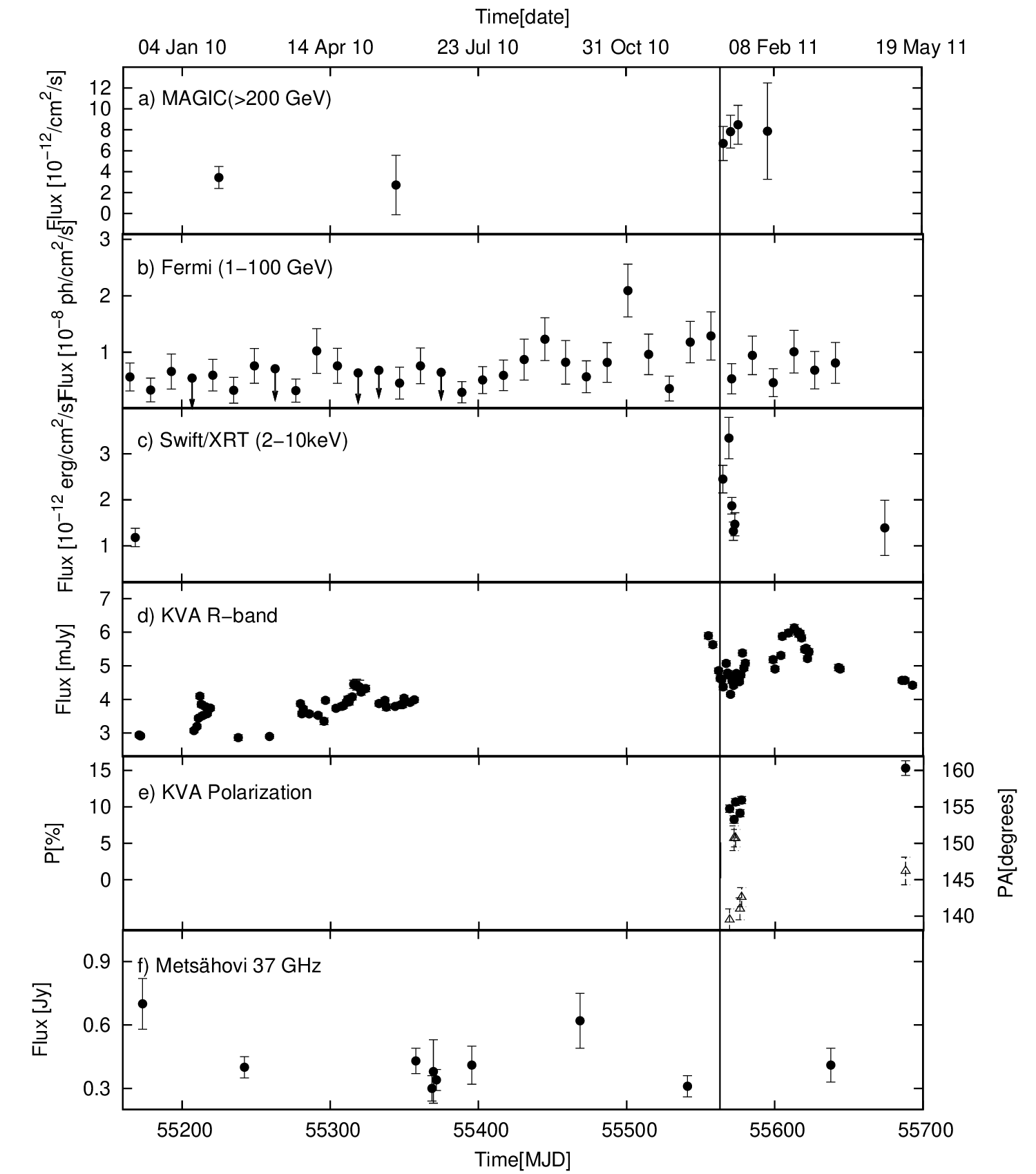}
\caption{Long-term multi-wavelength light curve of 1ES~1215+303 from
2009 December to 2011 May. The vertical line shows the beginning of
the MAGIC 2011 observation campaign. {\it a)} In the MAGIC light curve
2011 data are
  binned in 5-day intervals. 2010 data are divided in
  January-February and May-June bins. {\it b)} The
  {\it Fermi}-LAT light curve ($1-100$GeV) has bins of 14 days and the
points with
  arrows are upper limits. {\it c)} The {\it Swift}/XRT
  light curve is derived from the target of opportunity observations
  performed during the MAGIC observations and archival data. {\it d)}
The R-band light curve shows hourly average
  flux of the source, the error bars are smaller than
  the symbols in most cases. {\it e)} The optical polarization
   (filled circles, left axis) and polarization position angle
  (triangles, right axis) are hourly averages. {\it f)} 37\,GHz
  radio light curve from the Mets\"ahovi radio observatory.}
\label{fig_lc}
\end{figure}

\begin{figure}
\includegraphics[width=0.55\textwidth, angle=270]{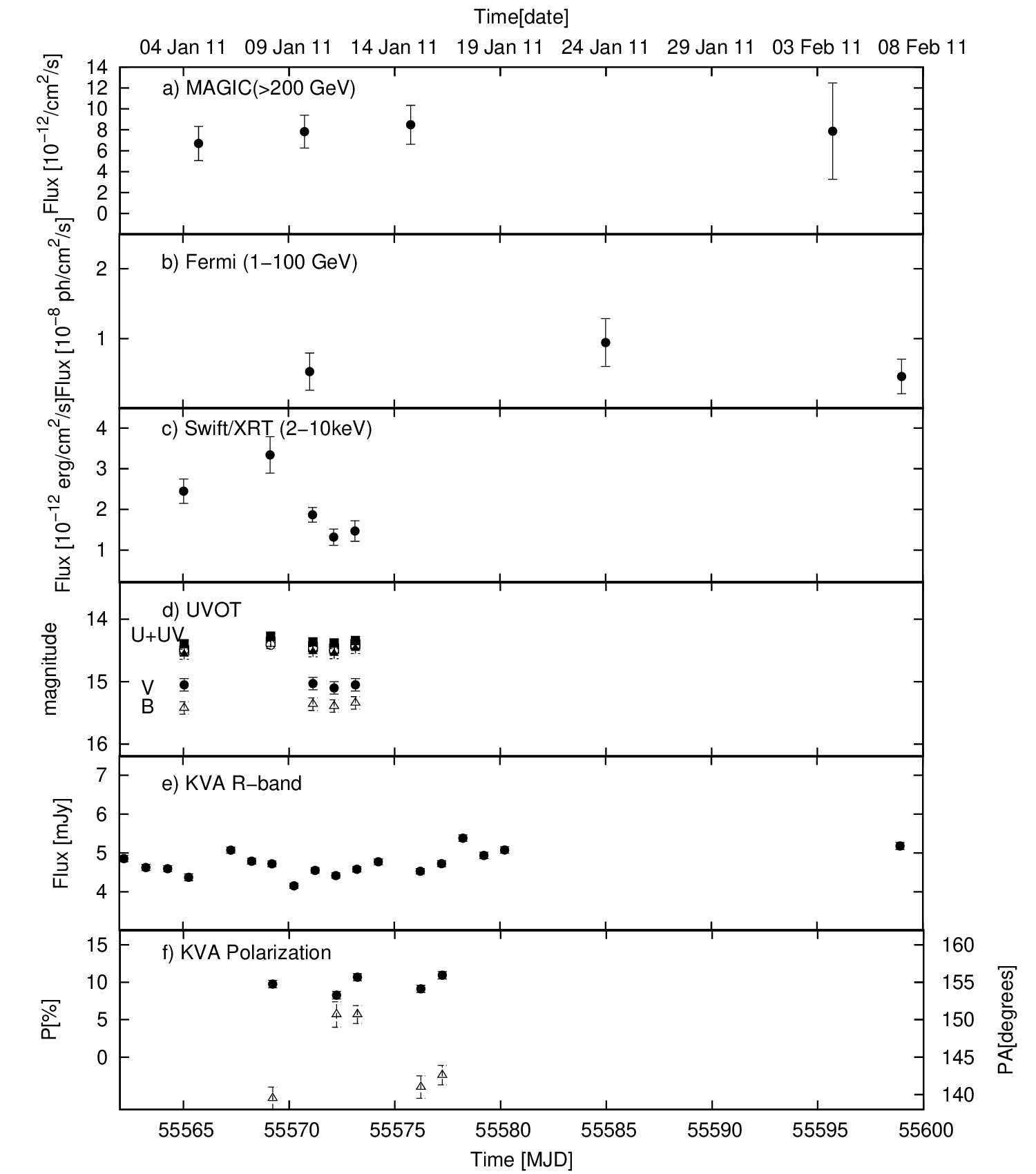}
\caption{Multi-wavelength light curve of 1ES~1215+303 from
  2011 January to February. {\it a)} In the MAGIC light curve, the data are binned in 5-day intervals. {\it b)} The
  {\it Fermi}-LAT light curve ($1-100$\,GeV) has bins of 14 days. {\it
c)} {\it Swift}/XRT
  light curve. {\it d)} UVOT optical and UV light curves.
  {\it e)} The R-band light curve shows hourly average
  fluxes of the source, the error bars are smaller than
  the symbols in most cases. {\it f)} The optical polarization
   (filled circles, left axis) and polarization position angle
  (triangles, right axis) are hourly averages. }
\label{fig_lc_short}
\end{figure}

The long-term multi-wavelength light curve, from radio to VHE
$\gamma$-rays, is shown in Fig.~\ref{fig_lc}. The MAGIC light curve
shows a lower flux in 2010 (January-February and May-June) than in 2011
(by a factor of 2). The large uncertainties in the {\it Fermi}-LAT
measurement do not allow us to conclusively say whether a similar flux
enhancement also occurred in the $1-100$\,GeV energy range (see section
3.2). In X-rays the source was in a high state (enhanced by a factor
of 2) in 2011 January compared with previous observations. In the
optical band the average flux during MAGIC observations in 2010 was 
3.5\,mJy, while in 2011 it was 4.6\,mJy. Thus, the source was clearly 
in outburst during early 2011, at least in VHE $\gamma$-rays, X-rays,
and the optical band. There were no simultaneous radio observations, but
both the previous and subsequent observations showed low flux,
suggesting that the outburst might have originated rather close
to the central engine where the emission region is opaque at radio
wavelengths. However, as the simultaneous observations are missing
the existence of a simultaneous radio flare cannot be excluded.
 
During the 2011 January-February observations
(Fig.~\ref{fig_lc_short}), the MAGIC light curve is consistent with a
constant flux. The source was in a rather low state in the {\it
Fermi}-LAT energy
range and no short term variability was detected. In X-rays and optical
the
source was variable during the MAGIC observations: the first two
X-ray exposures gave a higher flux than for the latter three. The X-ray
spectra show hints of hardening with higher flux, but they are
statistically the same.
The MAGIC observations started when the optical flux was decreasing,
but during 2011 January the optical light curve showed several small
flares. The X-ray light curve was more sparse and showed only one
flare, but the comparison of simultaneous optical and X-ray points
shows the same pattern in the light curves, indicating that the X-ray and optical emissions originates from the same region.


In addition to multi-wavelength variability studies, the optical
polarization measurements have proven to be a powerful tool to analyze
the emission scenarios in the blazar jets
\citep[e.g.][]{2008Natur.452..966M}. Polarization traces the
magnetic field of the jet. A net polarization oriented either parallel
or perpendicular to the projected jet axis can be confused by shocks 
and the signatures are visible in optical polarization. The
optical polarization measurements from 2011 January show little
variability in polarization degree (average $\sim 9\%$) or
PA (varying between $\sim 140^\circ$~-~$150^\circ$)
during the MAGIC observations, but the follow-up observations from
2011 April (Fig.~\ref{fig_lc}) show a higher polarization, 
$\sim 15\%$. Unfortunately, the polarization observations missed the
peak of the first optical outburst and our data sample is very
small. \cite{2011PASJ...63..639I} monitored the photo-polarimetric
behavior of the source in 2008-2009 and their observations seem to
show similar polarization trends (i.e. a decreasing polarization
 during outbursts). They also found that the PA was
almost constant at $\sim 150^\circ$, which agrees with our
observations and with the historical data from 1981-1989
\citep{2011ApJS..194...19W} showing PA values from $\sim
130-170^\circ$. Such preferred position angles have been
observed for several BL Lac objects
\citep[e.g.][]{1994ApJ...428..130J} and implies long-term stability of
the structure of the region producing the polarized emission
e.g. the existence of a optical polarization core. In first order, if
the optical
outburst was produced by a shock traveling along the jet, one would
expect the polarization degree to increase during the
outburst. However, if there is a standing shock (optical polarization
core) present, another shock with a different magnetic field
orientation
colliding with the standing component could produce an outburst in the
total flux, but decrease the observed level of polarization 
\citep{2010MNRAS.402.2087V}. A detailed photo-polarimetric study based
on more data would be needed to further test this hypothesis.

\subsection{Spectral energy distribution}

The SED of 1ES~1215+303 in both MAGIC observation epochs is shown in
Fig.~8. The 2011 high energy bump is constructed using the MAGIC
deabsorbed spectrum \citep[using the EBL model
of][]{2011MNRAS.410.2556D} and the simultaneous {\it Fermi}-LAT
spectrum (collecting
all photons from 2011 January-February). As stated in section 4.1, the
low energy bump was variable during the period and is constructed for
the night MJD 55569 that showed the highest {\it Swift} flux and for
which there are 
simultaneous KVA and UVOT observations. The contribution of the host
galaxy was subtracted from the R-band flux following
\cite{2007A&A...475..199N}. The host galaxy also contaminates the V, B,
and U bands of the UVOT data, but its contribution should be
negligible in the UV. As we have no direct measurements of the
host galaxy contribution in V, B and U bands we extrapolated the
magnitudes from the R-band value using the galaxy colors of elliptical
galaxies at $z=0.2$ \citep{1995PASP..107..945F}.

For the 2010 MAGIC data set, we could not derive a spectrum because of
the low significance of the signal 
but we report the flux between 300\,GeV and
1\,TeV (assuming the same spectral index as in 2011). The simultaneous
{\it Fermi}-LAT spectrum was calculated for the whole interval from 
2010 January to June. There was no simultaneous X-ray observation,
while for the optical we use the average (host galaxy subtracted) flux
from nights when MAGIC was also observing. This ``low state SED'' is
presented for illustrative purposes only but was not modeled, since both the
synchrotron and IC peaks are poorly constrained.

The SED of 2011 shows two peaks, with the synchrotron peak
frequency slightly above the optical band, as found for many other VHE
$\gamma$-ray emitting BL~Lac objects. The X-ray spectral index is also
typical for a BL Lac source. The second peak seems to be located
between
the {\it Fermi}-LAT and MAGIC points ($\sim$1\,GeV) as for many of
the VHE $\gamma$-ray emitting BL~Lacs. 
The locations of the synchrotron and IC peaks agree with values derived
in \cite{2010ApJ...716...30A}
for this source, but the synchrotron peak luminosity was slightly
higher than in the previous observation by \cite{2011arXiv1108.1114G}.

The emission characteristics of BL Lac objects is generally well
reproduced by the
one-zone leptonic model, in which a population of relativistic
electrons inside a region moving down the jet emit through synchrotron
and synchrotron self-Compton mechanisms
\citep{1996ApJ...461..657B,1998ApJ...509..608T}.
The spectral energy distribution in
2011 was modeled with the one-zone leptonic model fully described in
\cite{2003ApJ...593..667M}.
The emission region was assumed to be spherical, with
radius $R$, filled with a tangled magnetic field of intensity $B$ and
relativistic electrons, emitting synchrotron and synchrotron
self-Compton radiation. The electrons follow a smoothed broken power
law
energy distribution with normalization $K$ between $\gamma _{\rm min}$
and $\gamma _{\rm max}$, with slopes $n_1$ and $n_2$ below and above
the break at $\gamma _{\rm b}$. The relativistic boosting is fully
accounted for by the Doppler factor $\delta$. We note, however, that
one-zone models cannot reproduce the spectrum at the lowest frequencies,
since the emission is self-absorbed below the millimeter band. It is
generally assumed that this part of the SED is due to outer regions of
the jet that is not important for the modeling of the high-energy
emission.

\begin{figure}
\includegraphics[width=0.52\textwidth]{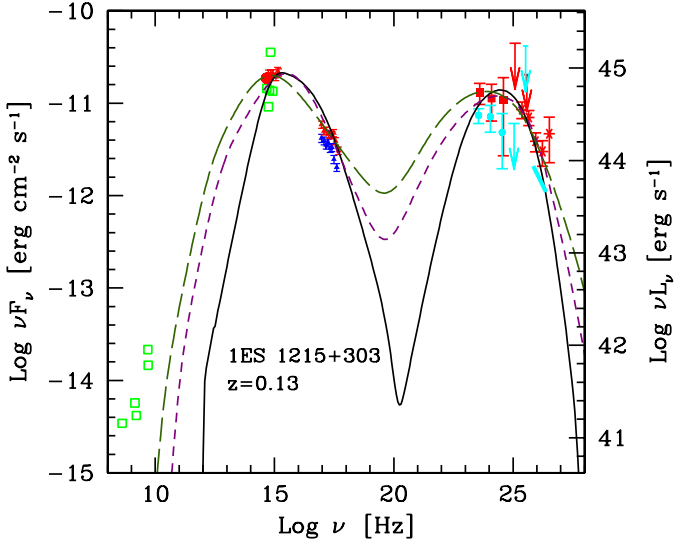}
\caption{Spectral energy distribution of 2011 January-February data
(red symbols) modeled with the one-zone SSC model of
\cite{2003ApJ...593..667M}.
From high to low energies: the deabsorbed MAGIC spectra (asterisk, see
text), the {\it Fermi}-LAT data (filled squares), Swift-XRT and
Swift-UVOT data (triangles: red for MJD 55569, blue for MJD 55565) and
simultaneous KVA data (filled circle, host galaxy subtracted, see
text). The cyan symbols report the January-June data of Fermi-LAT (data
points and arrows) and MAGIC (the thick oblique line).
{\it Fermi}-LAT (filled circles) and MAGIC (thick line) data. 
The green open squares are archival data. The dashed line is the model fit using
the extreme Doppler factor $\delta=60$, while the solid line is the
model fit with high $\gamma_{min}$ and the long dashed line reports the
model parameters that produces with smallest $\chi^2$ (see text and
Table 2).}
\label{fig_sed11}
\end{figure}

\begin{table*}[th]
\centering
\begin{tabular}{lcccccccccc}
\hline
\hline
model&$\gamma _{\rm min}$ & $\gamma _{\rm b}$ & $\gamma _{\rm max}$ &
$n_1$ & $n_2$ &$B$ & $K$ &$R$ & $\delta $ & $\chi^2/$d.o.f \\
&$[10^3$] & [$ 10^4$] &[$ 10^6$]  &  & &[G] & [cm$^{-3}]$  & $[10^{16}$
cm] & \\
\hline
high $\delta$ (dashed)&$1$ & $3$ & $1.0$ & $2.0$ & $4.2$ & $0.02$ &
$8\times 103$ & $0.8$ & $60$ & 3.36\\
high $\gamma_{min}$ (solid)&$8$ & $9.2$ & $2.5$ & $3.0$ & $4.85$ &
$0.055$ & $1.3\times 108$ & $1.0$ & $30$ & 6.94\\
min $\chi^2$ (long
dashed)&1&1.6&16.1&1.8&3.7&0.01&3.22$\times102$&3.75&36&1.04
\\
\hline
\hline
\end{tabular}
\vskip 0.4 true cm
\caption{Input model parameters for the three models shown in
Fig.\ref{fig_sed11}. 
The following quantities are reported: the minimum, break, and maximum
Lorentz factors and the low and high energy slope of the electron
energy distribution, 
the magnetic field intensity, the electron density, the radius of the
emitting region and its Doppler factor. In addition in the last column
we report the $\chi^2/$d.o.f assuming 2\%, 10\% and 40\% systematical
errors for optical-X-ray, GeV $\gamma$-rays and VHE $\gamma$-rays
respectively.}
\label{param}
\end{table*}

The optical-UV and X-ray data define a narrow synchrotron component
peaking around $10^{15}$ Hz. At high energies, the SSC bump is well
constrained by the {\it Fermi}-LAT and MAGIC data to peak at about
10\,GeV.
This particular structure of the SED is not easy to reproduce. In
particular, the relatively wide separation between the two peaks
inevitably implies a large value of the Doppler factor if standard
parameters are used for the electron energy distribution
\citep[e.g.][]{2003ApJ...594L..27G,2008MNRAS.385L..98T}.
Our best attempt to reproduce the data in the standard
framework provides the parameters given in Table 2 and is displayed 
as the dashed line of Fig.\ref{fig_sed11}. As expected from the
discussion above,
we find a large Doppler factor, $\delta=60$,  well
above the typical range of Doppler factors obtained from the modeling
of the emission of similar sources \cite[e.g.][]{2010MNRAS.401.1570T}
and disagreeing with the lower values required by the FR\,I-BL\,Lac
{\footnote{Fanaroff-Riley I radio galaxies (FR I)}}
unification scheme \citep{1995PASP..107..803U}.
However, there is an alternate, viable way to reproduce the observed
SED
using smaller Doppler factors;  
assume a relatively large minimum
Lorentz factor of the emitting electrons, $\gamma _{\rm min}=8\times
10^3$. This, along with a steep high energy electron energy
distribution  ($n_2=4.85$), allows us to properly reproduce the narrow
synchrotron bump and to locate the SSC peak at high enough energies
using a moderately large boosting, $\delta=30$. This solution
resembles one discussed for the case of BL Lacs showing hard spectra
in the soft X-ray and TeV band
\citep{2005A&A...433..479K,2009MNRAS.399L..59T,2011A&A...534A.130K,2011ApJ...740...64L}.
Interestingly, such parameters (large
$\gamma _{\rm min}$, steep slope) are consistent with the
prediction of some
simulations of particle acceleration by relativistic shocks
\citep[e.g.][]{2003heba.conf..157V,2011ApJ...726...75S}.
For example, for a proton-electron composition, it is expected that the
electrons are heated when crossing the shock to a typical Lorentz
factor of
$\Gamma$=$\Gamma _{\rm rel}\,m_p/m_e$, where $m_p/m_e=1836$ is the
proton to
electron mass ratio and $\Gamma _{\rm rel}=2-3$ is the relative
Lorentz factor between the upstream and the downstream flows. From this
$\Gamma$ (that is equivalent to our parameter $\gamma _{\rm min}$),
electrons are subsequently accelerated, forming a non-thermal tail that
is 
well approximated by a steep ($n=3.5$) power law.

The goodness of the fit can be judged by eye or by
$\chi^2$-minimization procedure. For the fits presented above the "eye
estimate" was
used, as for the latter the systematic errors of the data from
different instruments are in the key role. However, we also tested the
automatic
$\chi^2$-minimization procedure of \cite{2011ApJ...733...14M} 
with estimated systematical errors of 2\%, 10\% and 40\% for
optical-X-ray, GeV $\gamma$-rays and VHE $\gamma$-rays
respectively. The $\gamma_{min}$ is fixed to same value as in our high
$\delta$ model ($10^3$) to allow easier comparison. The resulting
parameters are shown in Table 2 and the fit with long-dashed (dark
green) line in Fig.~8. The minimal $\chi^2$ fit results in lower
$\delta$, but in a high $\gamma_{max}$ and rather large emission
region radius $R$ compared to other fits, but still compatible with the
day scale variability observed in X rays
and optical.  

\section{Summary and Conclusions}

In this paper the first detection of VHE $\gamma$-rays from
1ES~1215+303, resulting from MAGIC observations triggered by an
optical outburst of the source in 2011 January, has been reported. In
those data, the source is clearly detected at a 9.4\,$\sigma$
significance level. Also simultaneous
multi-wavelength data are presented from radio to HE $\gamma$-rays and
compared to
results from earlier MAGIC observations in 2010, when the source was
in a lower optical state. The VHE $\gamma$-ray flux in 2011 was higher
compared to 2010, suggesting that the activity in these two bands is
connected. This conclusion is further supported by the fact that
1ES~1215+303 is already the fifth discovery at VHE $\gamma$-rays
achieved after the MAGIC observations were triggered by an optical
outburst.

Our collected multi-wavelength data set is the most extensive energy
coverage for 1ES~1215+303 to date. The optical-VHE $\gamma$-ray
outburst seems to have been accompanied by an X-ray outburst, while in
the {\it Fermi}-LAT band the flux increased only marginally. The
optical photo-polarimetric data suggests that the high state could be
caused by a shock traveling down the jet that collides with a standing
shock with a differently oriented magnetic field. The X-ray and VHE
$\gamma$-ray high states could then also originate from this
collision.

The SED of 1ES~1215+303 in 2011 January
was modeled using a one-zone SSC model since it provides 
a good description of the SED of many VHE $\gamma$-ray emitting BL Lac
objects. However, for 1ES~1215+303 the synchrotron and IC
peaks are narrow, the separation between the two peaks is wide, and a
simple one-zone SSC model with typical
parameters failed to reproduce the observed SED. To fit
the SED, a high Doppler factor or a narrow electron energy distribution
is required. While high Doppler factors are disfavored by the
unified models, the high $\gamma_{min}$ value could be a viable
solution in the light of simulations modeling the acceleration of
electrons in a relativistic shock in a proton-electron jet. 
This should be further investigated, e.g. using the fully
self-consistent SSC model with particle acceleration due to shock and
stochastic acceleration by
\cite{2010ASTRA...6....1W,2010A&A...515A..18W}. 

Given the rather extreme conditions needed for the one-zone model, the
presence of a velocity structure in the jet
\citep{2003ApJ...594L..27G,2005A&A...432..401G} is also possible
and should be tested for modeling the SED. The narrow synchrotron and
IC peaks should well constrain the model. 
Additionally, the more complex emission scenario suggested by the
photo-polarimetric behavior of 1ES~1215+303 should be tested using, for
instance, a model of \cite{2011AAS...21832707M}, who investigated the
emission from a turbulent ambient jet plasma that passes through either
the standing or moving shocks in the jet.
Further observations of this exceptional VHE $\gamma$-ray
emitting BL Lac are strongly encouraged.

\section{Acknowledgments}

We would like to thank the Instituto de Astrof\'{\i}sica de
Canarias for the excellent working conditions at the
Observatorio del Roque de los Muchachos in La Palma.
The support of the German BMBF and MPG, the Italian INFN, 
the Swiss National Fund SNF, and the Spanish MICINN is 
gratefully acknowledged. This work was also supported by 
the Marie Curie program, by the CPAN CSD2007-00042 and MultiDark
CSD2009-00064 projects of the Spanish Consolider-Ingenio 2010
programme, by grant DO02-353 of the Bulgarian NSF, by grant 127740 of 
the Academy of Finland, 
by the YIP of the Helmholtz Gemeinschaft, 
by the DFG Cluster of Excellence ``Origin and Structure of the 
Universe'', by the DFG Collaborative Research Centers SFB823/C4 and
SFB876/C3,
and by the Polish MNiSzW grant 745/N-HESS-MAGIC/2010/0. The Mets\"ahovi
team acknowledges the support from the Academy of Finland
to our observing projects (numbers 212656, 210338, 121148, and others).

The \textit{Fermi}-LAT Collaboration acknowledges generous ongoing
support
from a number of agencies and institutes that have supported both the
development and the operation of the LAT as well as scientific data
analysis.
These include the National Aeronautics and Space Administration and the
Department of Energy in the United States, the Commissariat \`a
l'Energie Atomique
and the Centre National de la Recherche Scientifique / Institut
National de Physique
Nucl\'eaire et de Physique des Particules in France, the Agenzia
Spaziale Italiana
and the Istituto Nazionale di Fisica Nucleare in Italy, the Ministry of
Education,
Culture, Sports, Science and Technology (MEXT), High Energy Accelerator
Research
Organization (KEK) and Japan Aerospace Exploration Agency (JAXA) in
Japan, and
the K.~A.~Wallenberg Foundation, the Swedish Research Council and the
Swedish National Space Board in Sweden.

Additional support for science analysis during the operations phase is
gratefully acknowledged from the Instituto Nazionale di Astrofisica in
Italy and the Centre National d'\'Etudes Spatiales in France.

\bibliography{on325}

\end{document}